\documentclass[%
reprint,
 amsmath,amssymb,
 aps,
]{revtex4-2}

\usepackage{graphicx}% Include figure files
\usepackage{dcolumn}% Align table columns on decimal point
\usepackage{bm}% bold math
\usepackage[caption=false]{subfig}
\usepackage{siunitx, soul, xcolor}
\usepackage{adjustbox}

\begin{document}
\preprint{APS/123-QED}

\title{Atomic Layer Epitaxy of Kagome Magnet Fe$_3$Sn$_2$ and Sn-modulated Heterostructures}

\author{Shuyu Cheng}
\affiliation{Department of Physics, The Ohio State University, Columbus, Ohio 43210, United States}
\author{Igor Lyalin}
\affiliation{Department of Physics, The Ohio State University, Columbus, Ohio 43210, United States}
\author{Alexander J. Bishop}
\affiliation{Department of Physics, The Ohio State University, Columbus, Ohio 43210, United States}
\author{Roland K. Kawakami}
\email{kawakami.15@osu.edu}
\affiliation{Department of Physics, The Ohio State University, Columbus, Ohio 43210, United States}

\author{Binbin Wang}
\affiliation{Department of Materials Science and Engineering, The Ohio State University, Columbus, Ohio 43210, United States}
\author{N\'uria Bagu\'es}
\affiliation{Department of Materials Science and Engineering, The Ohio State University, Columbus, Ohio 43210, United States}
\author{David W. McComb}
\email{mccomb.29@osu.edu}
\affiliation{Department of Materials Science and Engineering, The Ohio State University, Columbus, Ohio 43210, United States}

\begin{abstract}
Magnetic materials with kagome crystal structure exhibit rich physics such as frustrated magnetism, skyrmion formation, 
topological flat bands, and Dirac/Weyl points. Until recently, most studies on kagome magnets have been performed on bulk crystals or polycrystalline films. 
Here we report the atomic layer molecular beam epitaxy synthesis of high-quality thin films of topological kagome magnet Fe$_3$Sn$_2$. 
Structural and magnetic characterization of Fe$_3$Sn$_2$ on epitaxial Pt(111) identifies highly ordered films with c-plane orientation and an in-plane magnetic easy axis.
Studies of the local magnetic structure by anomalous Nernst effect imaging reveals in-plane oriented micrometer size domains.
Superlattice structures consisting of Fe$_3$Sn$_2$ and Fe$_3$Sn are also synthesized by atomic layer molecular beam epitaxy, demonstrating the ability to modulate the sample structure at the atomic level.
The realization of high-quality films by atomic layer molecular beam epitaxy opens the door to explore the rich physics of this system and investigate novel spintronic phenomena by interfacing Fe$_3$Sn$_2$ with other materials.

\end{abstract}

\flushbottom
\maketitle
%  Click the title above to edit the author information and abstract
\thispagestyle{empty}

In recent years, studies on magnetic topological materials with kagome lattices have become one of the hottest frontiers of condensed matter research, due to their exotic physical properties in both real space and momentum space~\cite{smejkal2018,yang2017topological}. 
In momentum space, angle-resolved photoemission spectroscopy (ARPES) experiments on Mn$_3$Sn, Fe$_3$Sn$_2$, FeSn, and CoSn~\cite{kuroda2017,ye2018,kang2020,kang2020topological} show that kagome lattices give rise to Dirac cones and flat bands which are topologically protected and are of particular interest.
In addition, scanning tunneling spectroscopy finds evidence for topological flat bands as a sharp peak in the local density of states~\cite{yin2019negative}.
These topologically nontrivial features result in signatures of anomalous transport (e.g. chiral anomaly) in magnetotransport experiments~\cite{kuroda2017,chen2021}.
Furthermore, it is theoretically predicted that the band structures of the kagome topological magnets can be controlled by tuning of their magnetic structures~\cite{kuroda2017,smejkal2018}.
In real space, the kagome topological magnets have layered structures with spins occupying corner-sharing triangular lattices, which leads to geometrical spin frustration~\cite{fenner2009non,nakatsuji2015}.
A surprisingly large anomalous Hall effect (AHE) and magneto-optic Kerr effect (MOKE) have been reported in noncollinear antiferromagnet Mn$_3$Sn, even with vanishingly small net magnetization~\cite{nakatsuji2015,higo2018}.
Skymrmion spin textures have been observed in ferromagnetic  Fe$_3$Sn$_2$ resulting from the competition of exchange, dipolar, and Zeeman energies~\cite{hou2017,hou2019}.
However, most of the studies on the kagome magnets have been done on bulk materials ~\cite{fenner2009non,kida2011,nakatsuji2015,nayak2016,hou2017,ye2018,higo2018,hou2019,kang2020} with a few papers reporting the growth and characterization of epitaxial films~\cite{markou2018,inoue2019,taylor2020,khadka2020,hong2020molecular}.
Looking forward, the heterostructures consisting of kagome magnets will be interesting for both fundamental research and applications, due to the possibility of tuning the magnetic and topological properties via interface interactions, epitaxial strain, and quantum confinement.
However, all the reported studies on epitaxial Fe$_3$Sn$_2$ thin films have been focusing on high temperature growth so far, which may not allow for the formation of well-defined heterostructures due to interdiffusion at elevated temperatures.
Therefore, lower temperature growth of Fe$_3$Sn$_2$ is desired for the future development of heterostructures and superlattices based on kagome magnets.

\begin{figure}
    \subfloat[\label{fig:Structure}]{
        \includegraphics[width=0.38\textwidth]{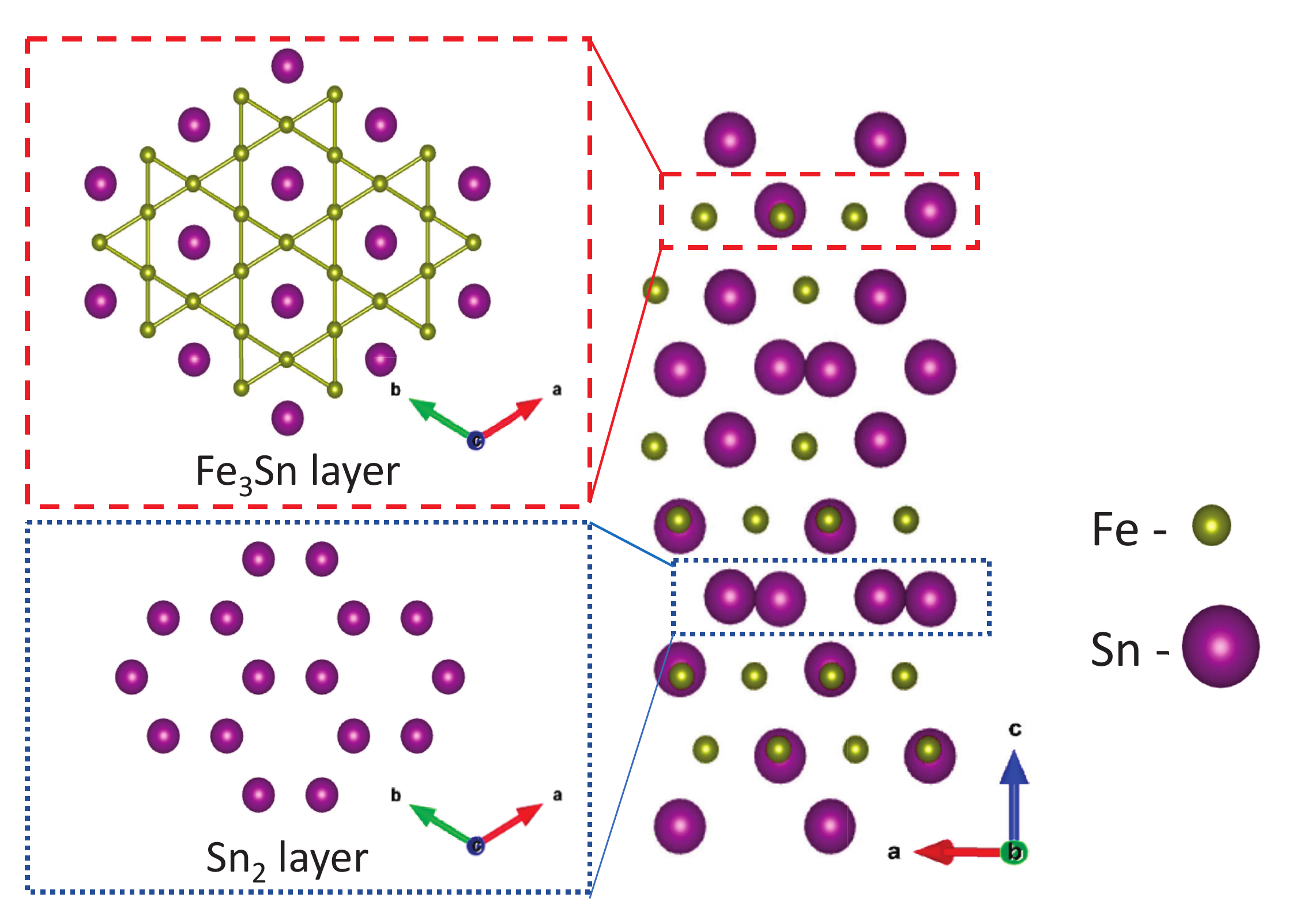}
    }
    \hfill
    \subfloat[\label{fig:RHEED}]{
        \includegraphics[width=0.45\textwidth]{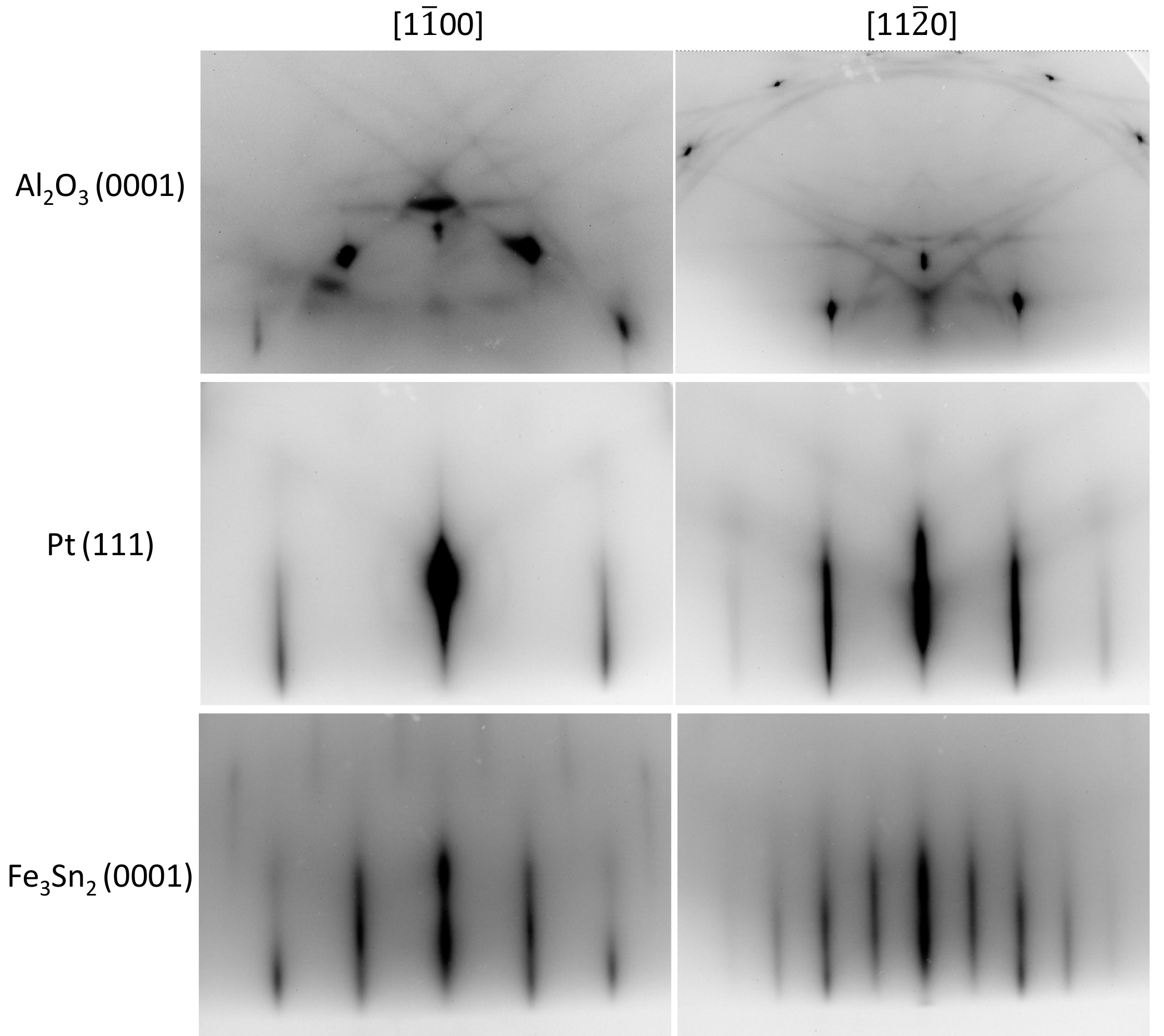}
    }
    \hfill
    \subfloat[\label{fig:RHEED_oscillations}]{
       \includegraphics[width=0.45\textwidth]{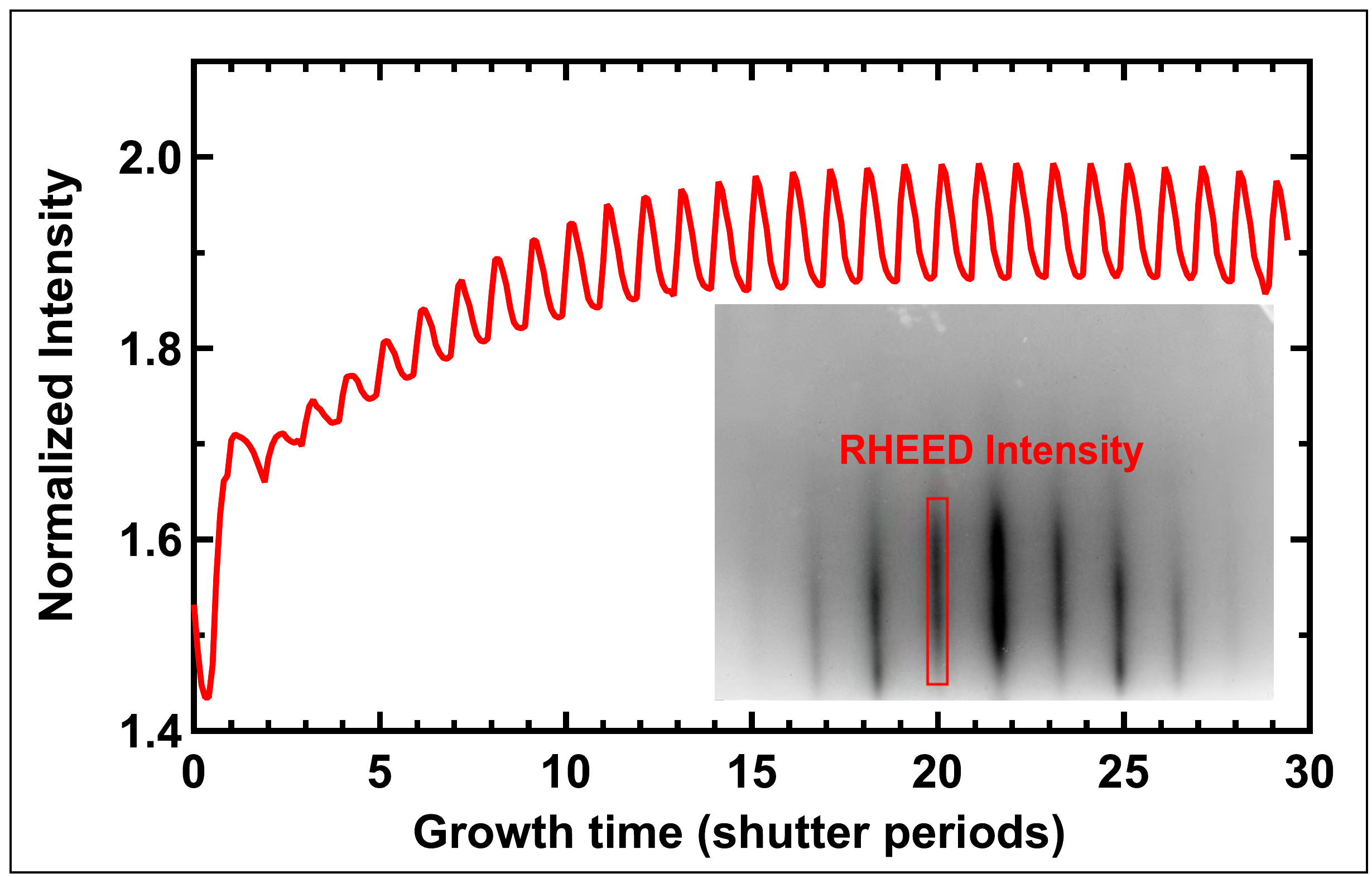}
    }
    \caption{Material structure and growth. 
    (a) Left: Top view of an individual Fe$_3$Sn layer with kagome structure (top) and Sn$_2$ layer with honeycomb structure (bottom), respectively. Right: Side view of the crystal structure of Fe$_3$Sn$_2$ consisting of alternating stacking of two Fe$_3$Sn kagome layers and one Sn$_2$ layer. 
    (b) RHEED patterns for the Al$_2$O$_3$(0001) substrate, 5\,nm Pt film, and 20\,nm Fe$_3$Sn$_2$ film measured with the beam along $[1\bar{1}00]$ (left column) and $[11\bar{2}0]$ (right column) directions of the substrate. 
    (c) Oscillations in the normalized RHEED intensity as a function of time. The RHEED intensity is measured within the red box and normalized by the intensity of the whole image as background.} 
\end{figure}

In this paper, we report the atomic layer molecular beam epitaxy (AL-MBE) growth of high-quality Fe$_3$Sn$_2$ thin films on Pt(111)/Al$_2$O$_3$(0001) substrates at lower temperatures.
By sequentially depositing Fe$_3$Sn kagome layers and Sn$_2$ layers (see Figure~\ref{fig:Structure}), we are able to control the sample structure at the atomic level.
The crystalline structure of our Fe$_3$Sn$_2$ sample is confirmed by a combination of \textit{in situ} reflection high energy electron diffraction (RHEED), X-ray diffraction (XRD) and transmission electron microscopy (TEM).
Energy-dispersive X-ray spectroscopy (EDX) shows sharp interfaces for low temperature growth.
The magnetic properties of Fe$_3$Sn$_2$ are investigated using MOKE, superconducting quantum interference device (SQUID) and anomalous Nernst effect (ANE).
Using a microscopy technique based on ANE, we successfully image the in-plane oriented domain structure of the epitaxial Fe$_3$Sn$_2$ films and investigate the magnetization reversal as a function of applied field.
We further utilize AL-MBE to precisely control the stacking sequences of Fe$_3$Sn and Sn$_2$ atomic layers, making superlattices with modulation of Sn$_2$ layers and confirm their structures by TEM and EDX.
This demonstrates the potential of using AL-MBE to generate designer materials consisting of kagome layers (Mn$_3$Sn, Fe$_3$Sn, Co$_3$Sn, etc.) and Sn$_2$ spacer layers with precision control of sample structures at the atomic level.

Fe$_3$Sn$_2$ is a ferromagnet with a high Curie temperature, $T_C$ = 670 K~\cite{giefers2006}, and saturation magnetization of 1.9\,$\mu_B$ per Fe at low temperature~\cite{ye2018}.
Fig.~\ref{fig:Structure} shows the crystal structure of Fe$_3$Sn$_2$ (space group R$\bar{3}$m, with lattice constants \textit{a} = 5.338\,\r{A} and \textit{c} = 19.789\,\r{A}~\cite{giefers2006}) which consists of Fe$_3$Sn kagome layers and Sn$_2$ spacer layers. 
In each Fe$_3$Sn monolayer, the Fe atoms form corner-sharing equilateral triangles surrounding hexagons, with Sn atoms sitting in the center of the hexagons.
The alternating sequence of one Sn$_2$ monolayer with honeycomb lattice and two Fe$_3$Sn kagome layers produces the layered crystal structure of Fe$_3$Sn$_2$.

Based on this layered structure, we synthesized Fe$_3$Sn$_2$ thin films on top of epitaxial Pt(111) buffer layers on Al$_2$O$_3$(0001) substates by AL-MBE.
The epitaxial growth was performed in an MBE chamber with a base pressure of $4\times10^{-10}$\,Torr. 
Films were deposited on Al$_2$O$_3$(0001) substrates (MTI Corporation) prepared by annealing in air at 1000\,$^\circ$C for 3 hours followed by annealing in ultrahigh vacuum (UHV) at 500\,$^\circ$C for 30 minutes. 
A 5\,nm Pt(111) buffer layer was deposited from an e-beam evaporator (Pt: 99.99\%, Kurt J. Lesker) onto the Al$_2$O$_3$ (0001) substrate by growing the first 0.6\,nm at 440\,$^\circ$C and the rest 4.4\,nm while cooling down from 140\,$^\circ$C to 80\,$^\circ$C. 
The Pt buffer layer was post-annealed at 300\,$^\circ$C to improve the crystallinity and surface roughness. 
The Fe$_3$Sn$_2$ layer was grown on Pt(111) at 100\,$^\circ$C using the following AL-MBE sequence: deposit two atomic layers of Fe$_3$Sn with a Fe:Sn flux ratio of 3:1, deposit one atomic layer of Sn$_2$ with the growth time same as two Fe$_3$Sn layers, then repeat. 
The Fe and Sn fluxes were generated from Knudsen cells (Fe: 99.99\%, Alfa Aesar; Sn: 99.998\%, Alfa Aesar) and the growth rates were determined using a quartz deposition monitor that was calibrated by x-ray reflectometry. 
Typical growth rates were $\sim$ 0.85\,\r{A}/min, $\sim$ 0.67\,\r{A}/min, and $\sim$ 0.45\,\r{A}/min for Fe, Sn, and Pt, respectively. 
To protect the sample from oxidation, a 5\,nm CaF$_2$ capping layer was deposited on top of the Fe$_3$Sn$_2$. 

RHEED patterns were measured during growth to characterize the epitaxial growth and determine the in-plane lattice constants. 
Figure~\ref{fig:RHEED} shows the RHEED patterns for the Al$_2$O$_3$(0001) substrate (top row), 5\,nm Pt buffer layer (middle row), and the Fe$_3$Sn$_2$ layer after 20\,nm of growth (bottom row). 
The left and right columns show patterns taken for the beam along the $[1\bar{1}00]$ and $[11\bar{2}0]$ directions of the substrate, respectively.
With in-plane rotation of the sample, RHEED exhibits the same pattern every 60$^\circ$ (i.e. six-fold rotation symmetry) with the patterns alternating between $[1\bar{1}00]$-type and $[11\bar{2}0]$-type every 30$^\circ$. 
For the in-plane epitaxial alignment, the Pt(111) and Fe$_3$Sn$_2$(0001) surface unit cells are aligned with each other and rotated 30$^\circ$ with respect to the Al$_2$O$_3$(0001) substrate, as $a_{Al_2O_3} \approx (\sqrt{3}/2)2a_{Pt} \approx (\sqrt{3}/2)a_{Fe_3Sn_2}$ (bulk values for in-plane lattice constant are $a_{Al_2O_3} = 4.759$\,\r{A}, $2a_{Pt} = 5.549$\,\r{A}, $a_{Fe_3Sn_2}=5.338$\,\r{A}).

The streaky patterns observed during Fe$_3$Sn$_2$ growth signify diffraction from a two-dimensional surface. 
In addition, we observe oscillations (Fig.~\ref{fig:RHEED_oscillations}) in the normalized RHEED intensity where the maxima occurs for the Fe$_3$Sn termination and the minima occurs for the Sn$_2$ termination. 
The normalization is performed by dividing the intensity of the background and is helpful for canceling variations in the incident beam intensity and background lighting. 
Except for the change in RHEED intensity, we did not observe any other significant differences in the RHEED pattern between Sn$_2$ and Fe$_3$Sn terminations. Nevertheless, the presence of RHEED oscillations in atomic layer MBE confirms the modulation of the surface termination during growth.

\begin{figure*}

\subfloat[\label{fig:XRD}]{
    \includegraphics[width=0.45\textwidth]{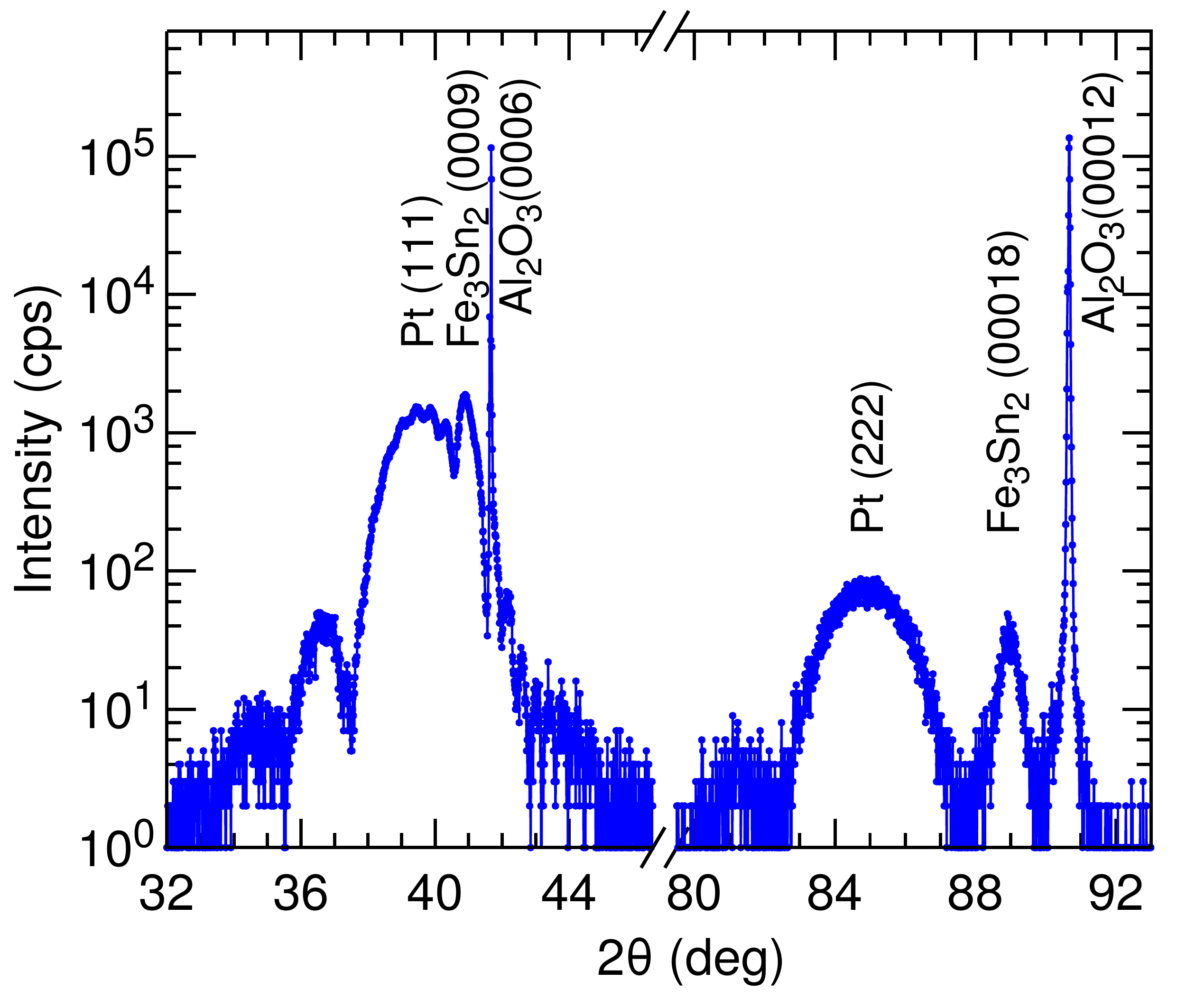}
    }
    \subfloat[\label{fig:AFM}]{
       \includegraphics[width=0.5\textwidth]{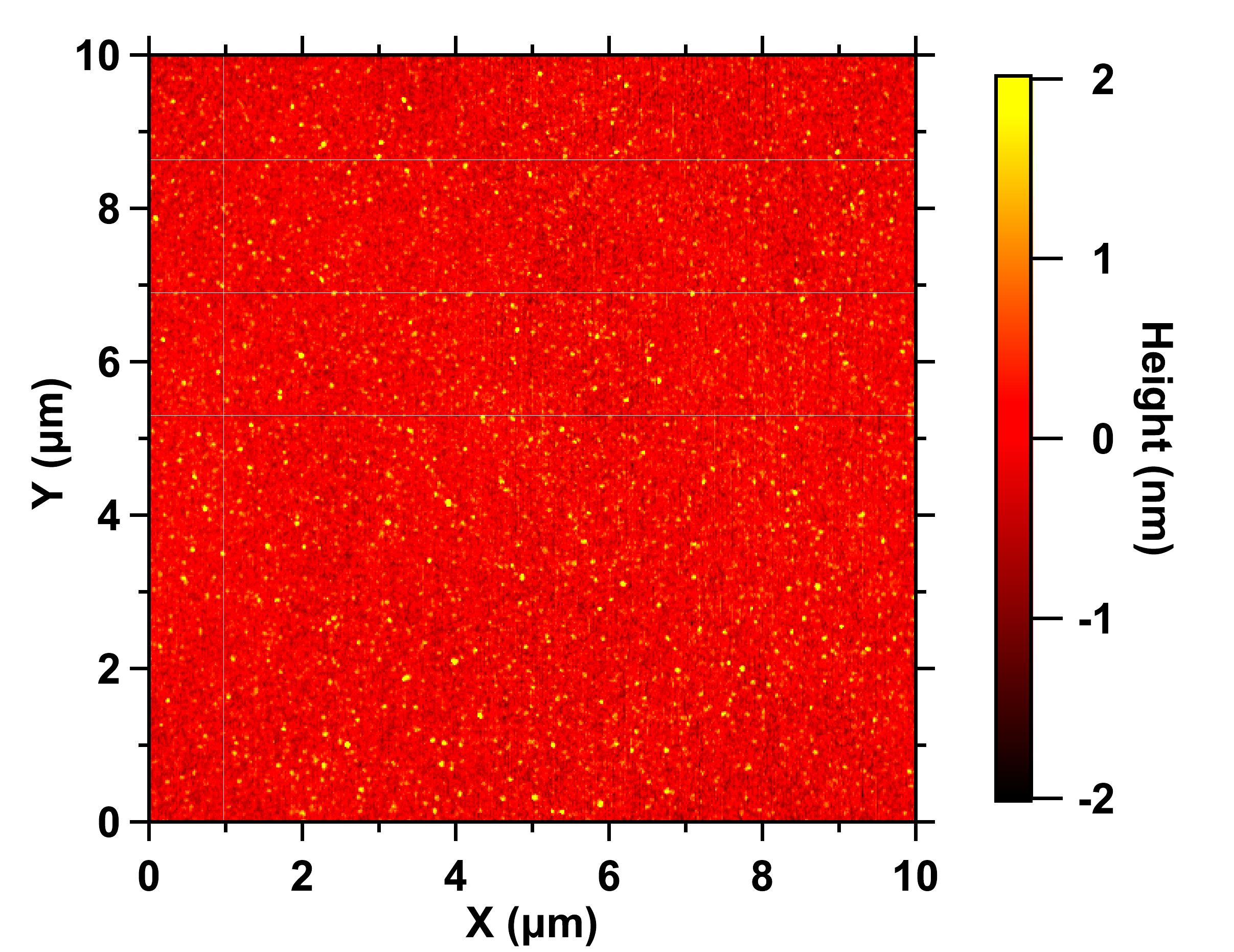}
    }
    \hfill
    \subfloat[\label{fig:EDX}]{
       \includegraphics[height = 3.2in]{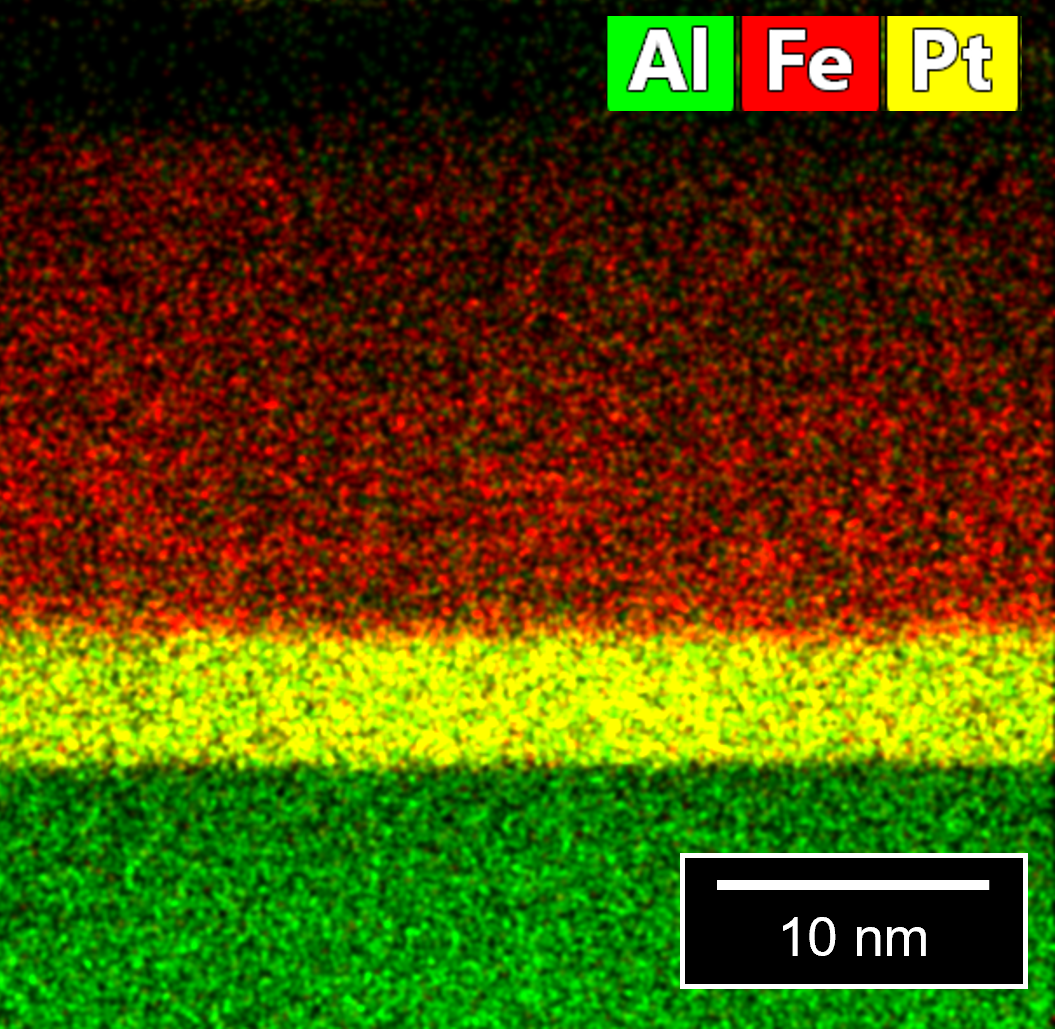}
    }
    \subfloat[\label{fig:TEM}]{
       \includegraphics[height = 3.2in]{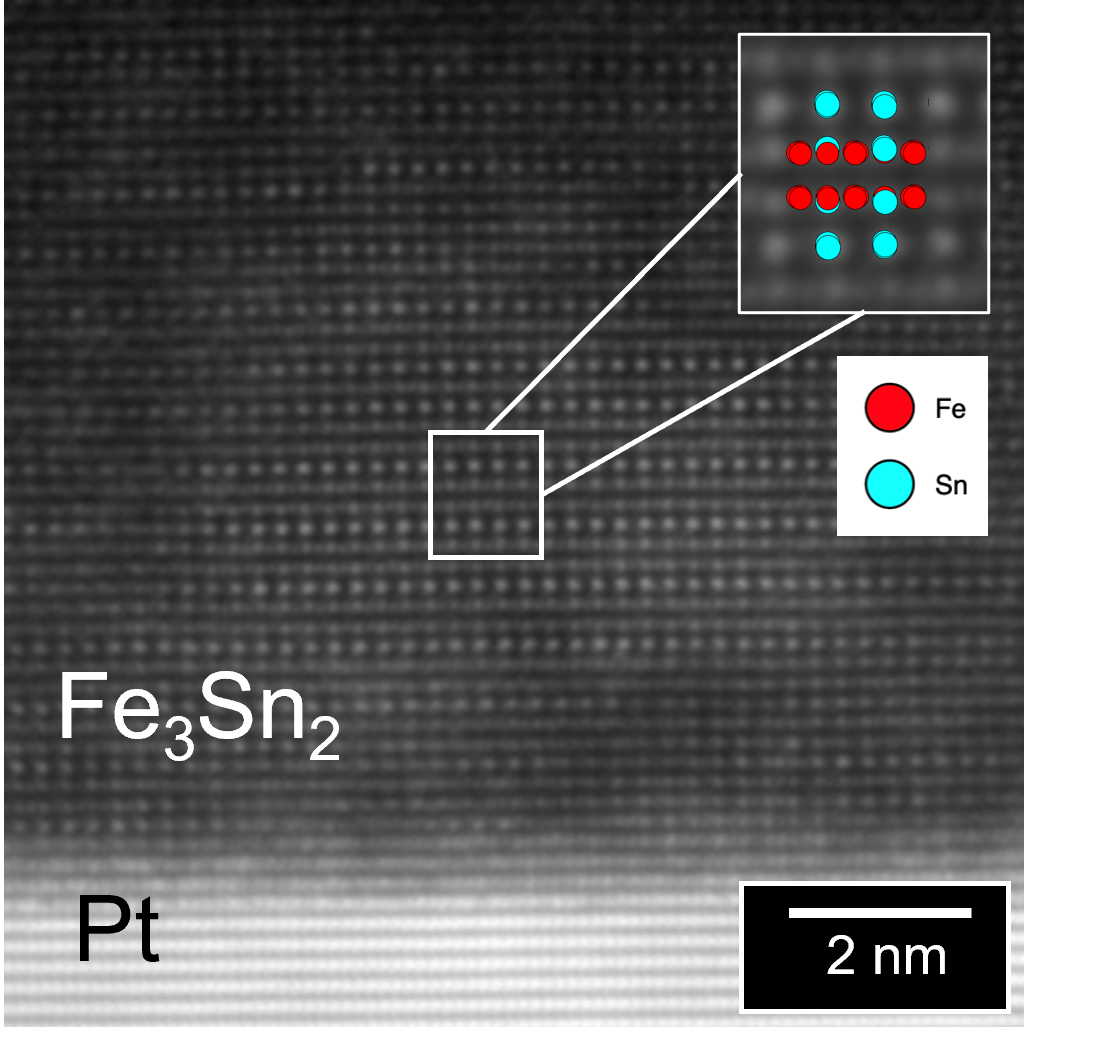}
    }
  \caption{Material characterization. 
  (a) 2$\theta$-$\omega$ scan of a 20\,nm Fe$_3$Sn$_2$ film grown on c-sapphire. 
  %(b) A triple axis $\omega$-relative scan (rocking curve) of the Fe$_3$Sn$_2$ (009) peak (red) with a Gaussian fitting (blue). 
  (b) AFM scan over a 10 $\mu$m $\times$ 10 $\mu$m area on the surface of a 20\,nm Fe$_3$Sn$_2$ film (rms roughness 0.362\,nm). 
  (c) Cross-sectional EDX-STEM chemical map of a 20\,nm Fe$_3$Sn$_2$ sample grown at 100\,$^\circ$C. 
  (d) Atomic resolution HAADF-STEM imaging of a 20\,nm Fe$_3$Sn$_2$ sample grown at 100\,$^\circ$C viewed along the Fe$_3$Sn$_2$ [11$\bar{2}$0] direction. The upper right insert is magnified from the white box and overlaps the projected unit cell of Fe3Sn2 along Fe$_3$Sn$_2$ [11$\bar{2}$0] direction.}
\end{figure*}

Films grown by this method were studied with XRD using Cu K-$\alpha$ line (wavelength 1.5406\,\r{A}) to analyze their crystal structure.
A representative $\omega$-$2\theta$ scan of a 20\,nm film grown at 100\,$^\circ$C is shown in Figure~\ref{fig:XRD} and includes the Fe$_3$Sn$_2$ (0009) peak with several Laue oscillations, indicating a smooth film.
The out-of-plane lattice parameter extracted from analysis of this scan is 19.85\,\r{A} which agrees well with previous reports of 19.789\,\r{A} \cite{giefers2006}.
A peak from the 5\,nm Pt(111) buffer also shows Laue oscillations with larger angular period due to smaller thickness of the Pt layer in comparison with Fe$_3$Sn$_2$ layer (Pt Laue peaks at $\sim$34.5$^\circ$, $\sim$36.7$^\circ$, and $\sim$43.5$^\circ$) demonstrating high quality of the buffer layer.
At larger $2\theta$ angles we observe Fe$_3$Sn$_2$ (00018) and Pt(222) peaks, with no additional peaks that could be attributed to impurity phases (see Supplementary Material (SM) sections 2 for full range scans).

To characterize the surface topography of the sample, we performed atomic force microscope (AFM) measurements on uncapped 20\,nm Fe$_3$Sn$_2$ films. 
Figure~\ref{fig:AFM} shows a typical 10 $\mu$m $\times$ 10 $\mu$m scan of a 20\,nm Fe$_3$Sn$_2$ sample grown at 100\,$^\circ$C. 
The AFM image indicates that the sample has flat surface, with root-mean-square (rms) roughness of 0.362\,nm.

An important factor for material synthesis is the growth temperature. To optimize the growth temperature, we performed AFM and XRD measurements on a series of samples grown at different temperatures ranging from room temperature to 200\,$^\circ$C. The AFM and XRD results are shown in SM sections 1 and 2, respectively. We conclude that 100\,$^\circ$C is the optimized growth temperature as it gives the best AFM roughness and sharp XRD peaks.

The epitaxial quality of the 20\,nm Fe$_3$Sn$_2$ sample grown at 100\,$^\circ$C was examined using a probe-corrected Themis Z S/TEM at 200 kV. 
Figures \ref{fig:EDX} and \ref{fig:TEM} show the  energy-dispersive x-ray (EDX) chemical map and cross-sectional scanning transmission electron microscopy (STEM) image, revealing a clear interface between the Pt buffer and Fe$_3$Sn$_2$ thin film.
The stoichiometry of Fe$_3$Sn$_2$ thin films was confirmed by electron energy loss spectroscopy (EELS), which gives an atomic ratio of Fe:Sn $\approx$ 1.5 (see SM sections 3 for details).
To identify the crystalline quality of the Fe$_3$Sn$_2$ thin films, atomic resolution high-angle annular dark-field (HAADF) STEM images of the Fe$_3$Sn$_2$/Pt interface were acquired along the Fe$_3$Sn$_2$ [11$\bar{2}$0] direction (see Figure~\ref{fig:TEM}). 
Since the contrast in HAADF STEM is approximately proportional to the square of the atomic number, the Sn atoms in the Sn$_2$ atomic layers appear as the brightest spots, while the atoms in Fe$_3$Sn kagome layers are dimmer (the atomic numbers of Sn and Fe are 50 and 26, respectively). 
The alternating sequence of one Sn$_2$ monolayer and two Fe$_3$Sn kagome layers shows a highly crystalline film with the expected Fe$_3$Sn$_2$ phase, although some stacking faults are observed.

\begin{figure*}
    \subfloat[\label{fig:MOKE}]{
   \includegraphics[width=0.5\textwidth]{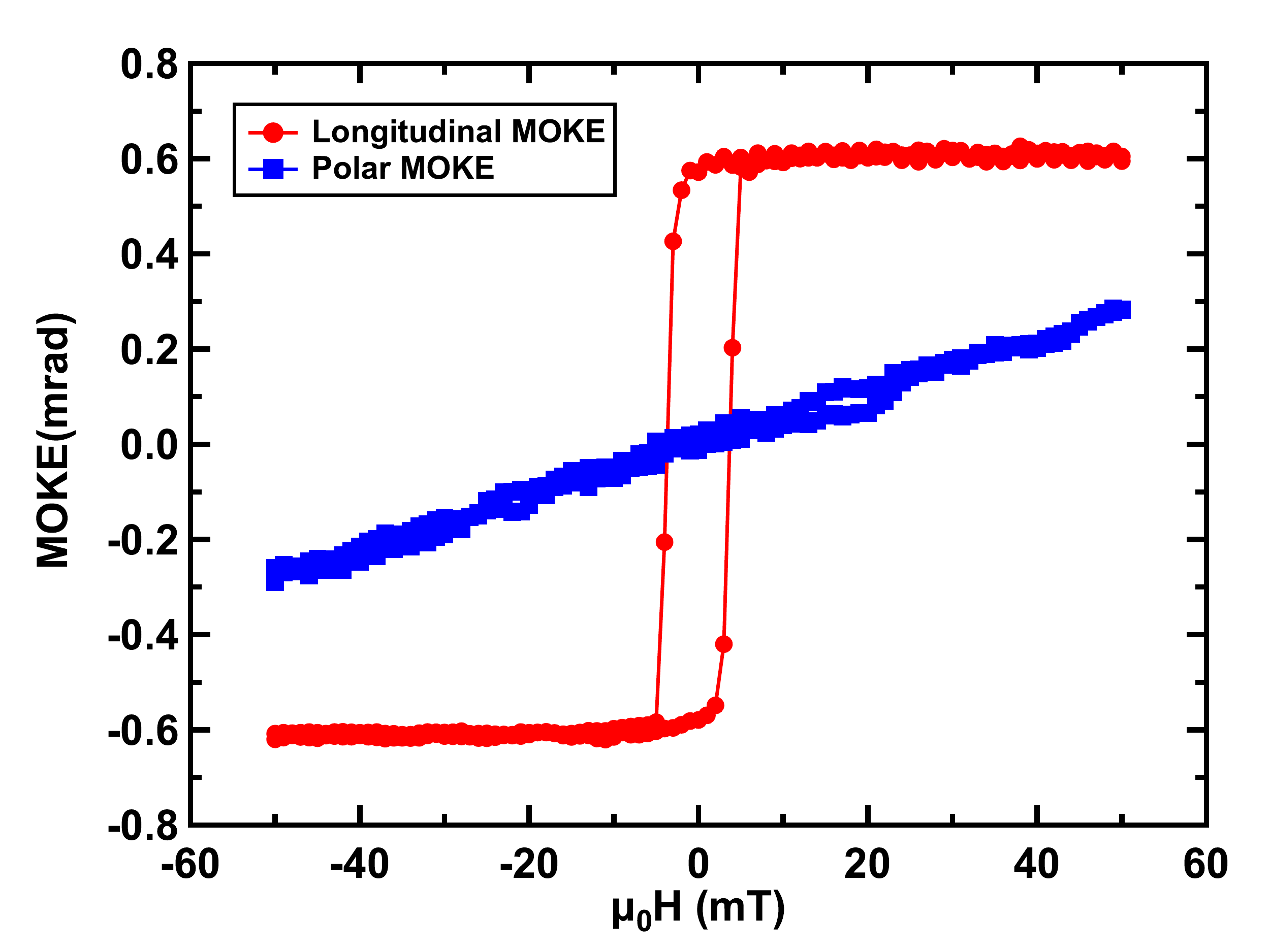}
}
   \subfloat[\label{fig:SQUID}]{
    \includegraphics[width=0.5\textwidth]{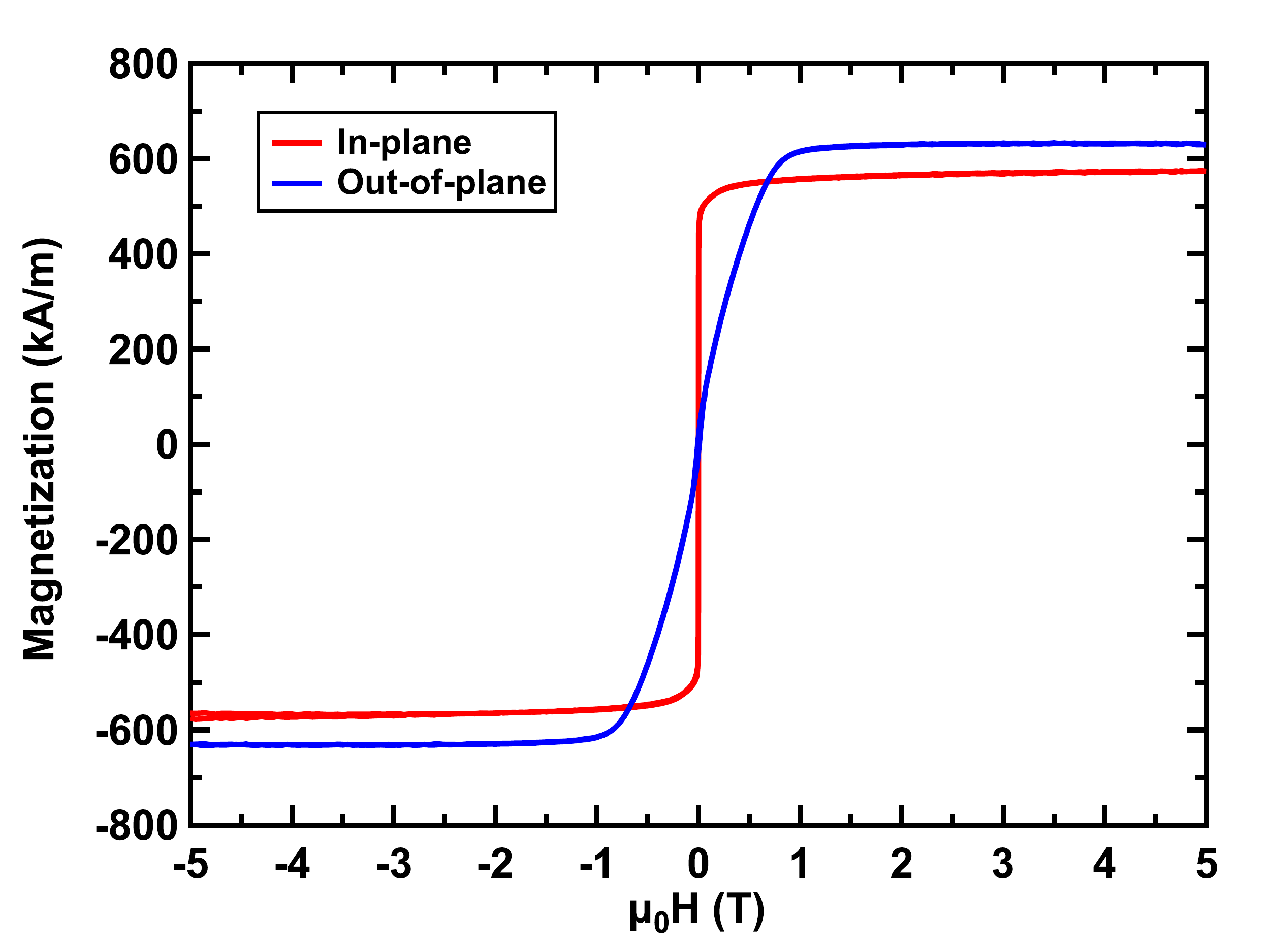}
    }
  \caption{Magnetic properties of Fe$_3$Sn$_2$ films. 
  (a). Longitudinal (red) and polar (blue) MOKE hysteresis loops of a 20\,nm Fe$_3$Sn$_2$ film. 
  (b). Magnetic hysteresis loops of a 20\,nm Fe$_3$Sn$_2$ film measured using SQUID magnetometry with in-plane (red) and out-of-plane (blue) geometries.}
\end{figure*}

To investigate the in-plane and out-of-plane magnetic properties of the Fe$_3$Sn$_2$ films, we measured longitudinal and polar MOKE hysteresis loops. 
The samples were probed using a linearly-polarized He-Ne laser (633\,nm wavelength, $\sim100~\mu$W power, $\sim100~\mu$m spot size) and a polarizing beamsplitter, photodiode bridge, and lock-in amplifier (463 Hz intensity modulation) to detect the Kerr rotation. 
The laser beam had a $\sim$45$^\circ$ angle of incidence for longitudinal MOKE and normal incidence for polar MOKE.
Figure~\ref{fig:MOKE} shows a representative longitudinal hysteresis loop (red curve) measured on a 20\,nm thick Fe$_3$Sn$_2$ sample. 
The square hysteresis loop with a coercivity of 2.4\,mT indicates ferromagnetic order with in-plane magnetization. 
In contrast, the polar hysteresis loop (blue curve) shows a small Kerr rotation with slight variation with out-of-plane magnetic field. 
We repeated the longitudinal MOKE measurements on additional samples with thicknesses varying from 5\,nm to 20\,nm. 
The coercive fields of the samples are between 3.0\,mT and 4.2\,mT (SM section 4).
Together, the longitudinal and polar MOKE loops show that the Fe$_3$Sn$_2$ samples have an easy-plane magnetic anisotropy.
This agrees with previous studies of Fe$_3$Sn$_2$ films grown by sputter deposition~\cite{khadka2020} and in bulk crystals thinned to below $\sim$100\,nm~\cite{wang2021stimulated}.

We also performed SQUID magnetometry measurements on our Fe$_3$Sn$_2$ films. 
Figure~\ref{fig:SQUID} shows hysteresis loops of a 20\,nm thick Fe$_3$Sn$_2$ sample measured with in-plane (red curve) and out-of-plane (blue curve) magnetic fields. 
The in-plane hysteresis loop exhibits a sharp switching behavior while the out-of-plane hysteresis loop exhibits almost linear behavior within $\pm$1\,Tesla and saturates at $\sim$1\,Tesla, suggesting that our Fe$_3$Sn$_2$ samples have easy-plane anisotropy. 
Furthermore, our SQUID results give a saturation magnetization $M_s$ = 630\,kA/m, which is consistent with previous studies~\cite{khadka2020, kida2011giant}.

\begin{figure*}
    \subfloat[\label{fig:ANE_setup}] {
    \includegraphics[width=0.40\textwidth]{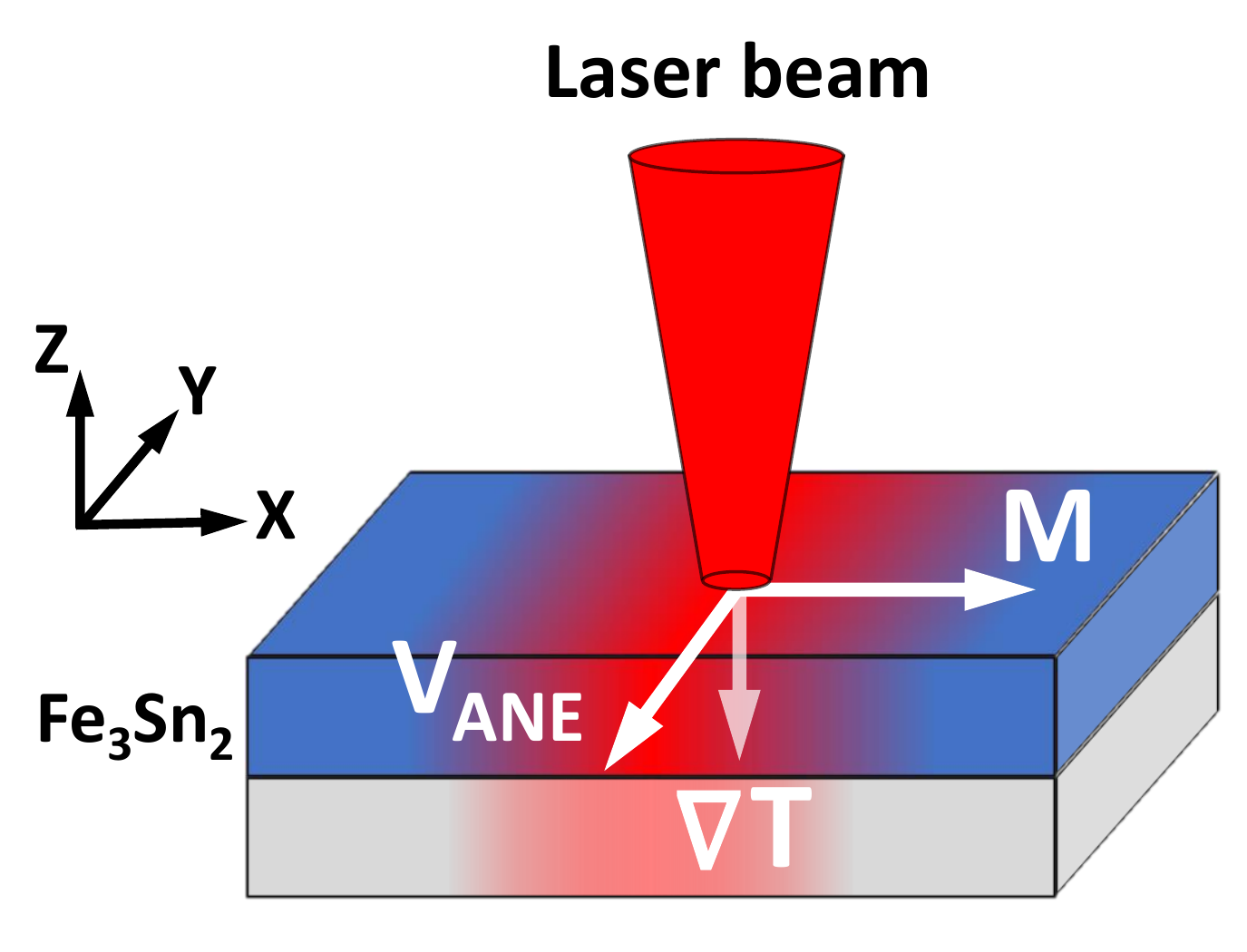}
    }
  \hfill
    \subfloat[\label{fig:ANE_device}] {
    \includegraphics[width=0.14\textwidth]{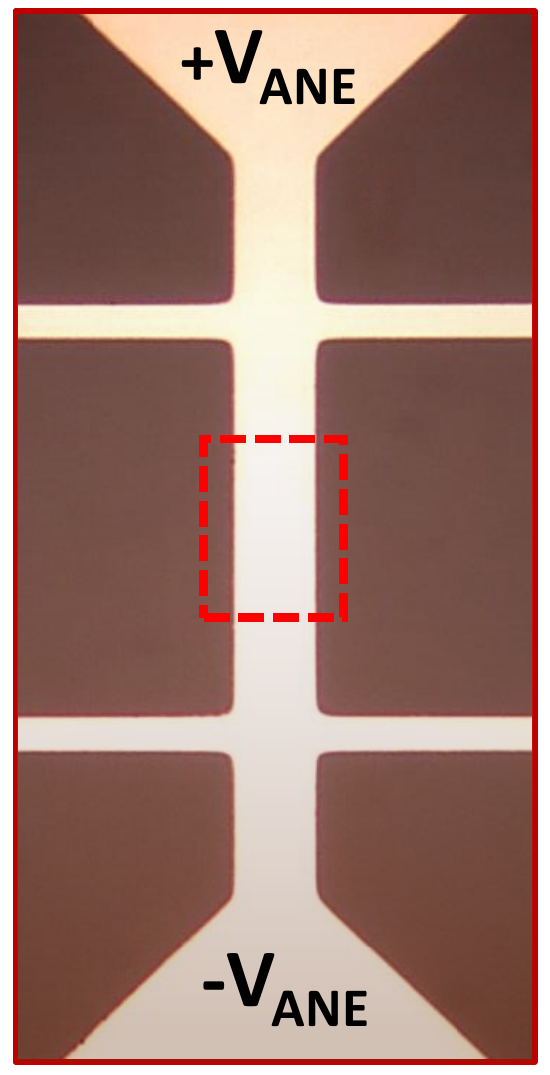}
    }
  \hfill
    \subfloat[\label{fig:ANE_hloop}]{
    \includegraphics[width=0.35\textwidth]{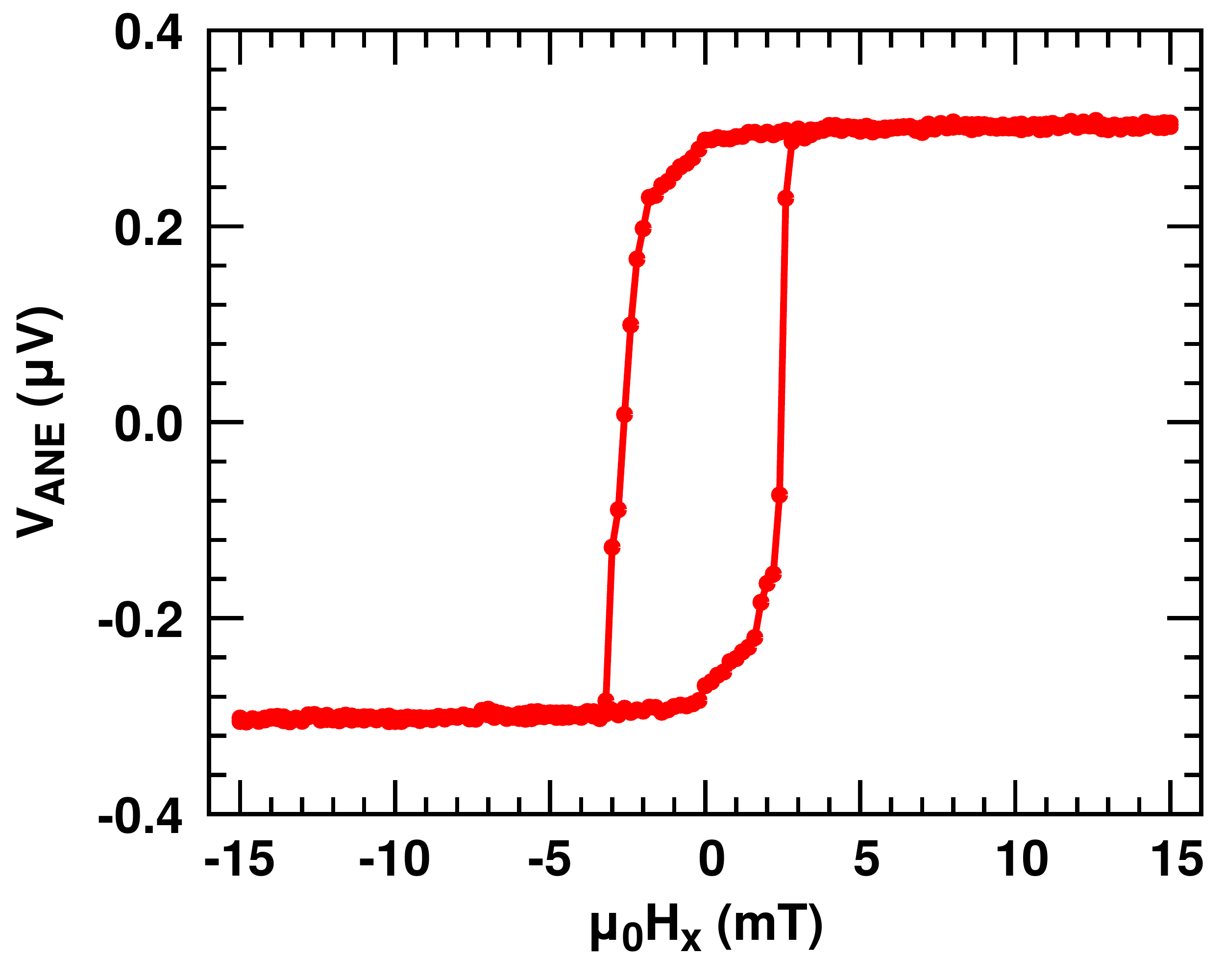}
    }
  \hfill
\subfloat[\label{fig:ANE_imaging}] {
    \includegraphics[width=\textwidth]{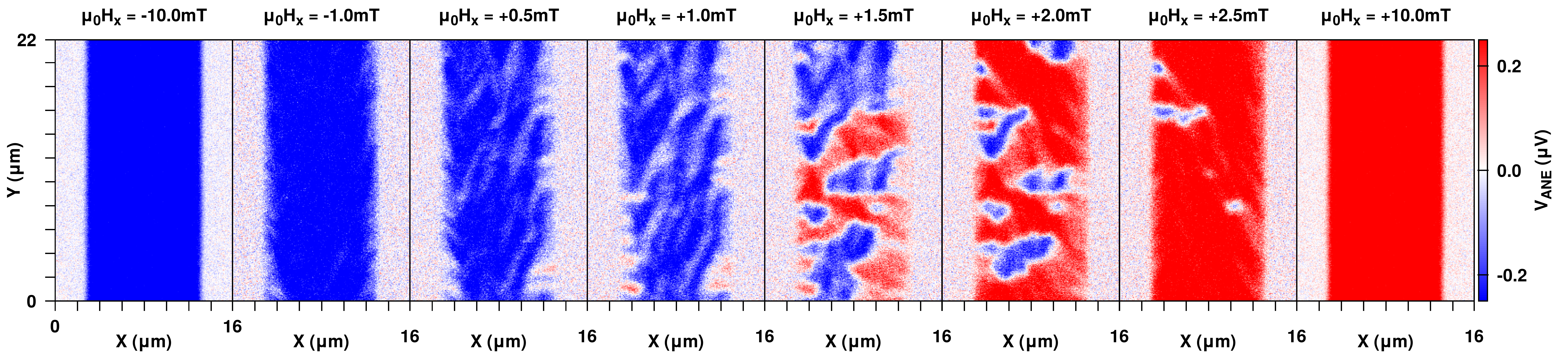}
    }
  \caption{Anomalous Nerst effect imaging.
  (a) Schematics of thermal gradient microscopy.
  The laser beam is scanned over the sample surface, the induced local ANE voltage reflects the local magnetic properties.
  (b) Microscope image of a typical device, (the dashed rectangle corresponds to the area imaged in (d))
  (c) ANE hysteresis loop of a 20\,nm Fe$_3$Sn$_2$ film.
  (d) Magnetization reversal of a 20\,nm film through multidomain state imaged by ANE at a series of magnetic fields $\mu_0$H$_x$.}
\end{figure*}

The magnetic domain structures of Fe$_3$Sn$_2$ films are of interest due to the observation of skyrmions in bulk Fe$_3$Sn$_2$, but has not yet been studied in thin films. 
Longitudinal MOKE microscopy with oblique
angle incidence can detect the in-plane magnetization and therefore determine in-plane domain structure of our Fe$_3$Sn$_2$ films. However, in this manuscript, we choose to use thermal gradient microscopy (TGM)~\cite{weiler2012,gray2019,reichlova2019} over longitudinal MOKE to image domain structure because we found that it has a better signal-to-noise ratio in our experimental setup.

TGM is based on moving a laser spot over the sample surface, and recording a voltage induced by the local laser heating.
The thermal gradient generated in the out-of-plane direction $Z$ and a component of magnetization in the $X$ direction give rise to the anomalous Nernst effect, which is detected as a voltage along the $Y$ direction, $V_{ANE} \sim \left[ \nabla T \times \mathbf{M} \right]$ (see Fig.~\ref{fig:ANE_setup}).

For the ANE imaging, we fabricated 10\,$\mu$m wide Hall bar devices by a combination of photolithography and argon ion milling (Fig.~\ref{fig:ANE_device}). The laser excitation for the thermal gradient was produced by a frequency-doubled
(BaB$_2$O$_4$ crystal) 
mode-locked Ti:Sapphire laser for a wavelength of 400\,nm. The laser beam with 0.7\,mW power was focused by a 50$\times$ objective lens (NA of 0.6) to a spot size of 0.9\,$\mu$m, and a fast steering mirror in the 4f alignment scheme was used for scanning the laser spot over the sample surface. The intensity of the beam was modulated at a frequency of 120 kHz and the generated ANE voltage was detected using a lock-in amplifier.

\begin{figure*}
   \subfloat[\label{fig:RHEED_superlattice}]{
   \includegraphics[height = 1.9in]{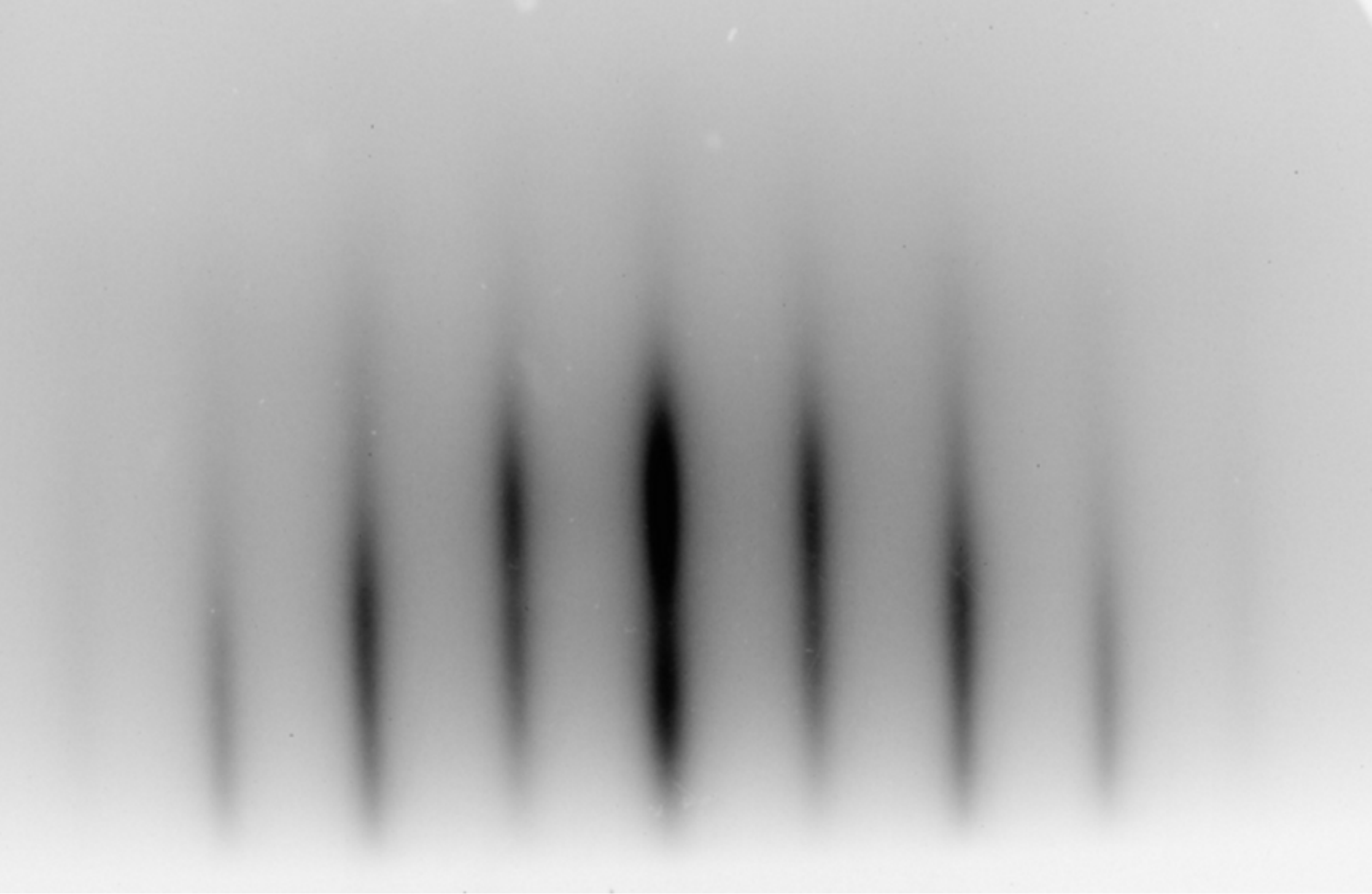}
}
   \subfloat[\label{fig:TEM_superlattice}]{
   \includegraphics[height = 1.9in]{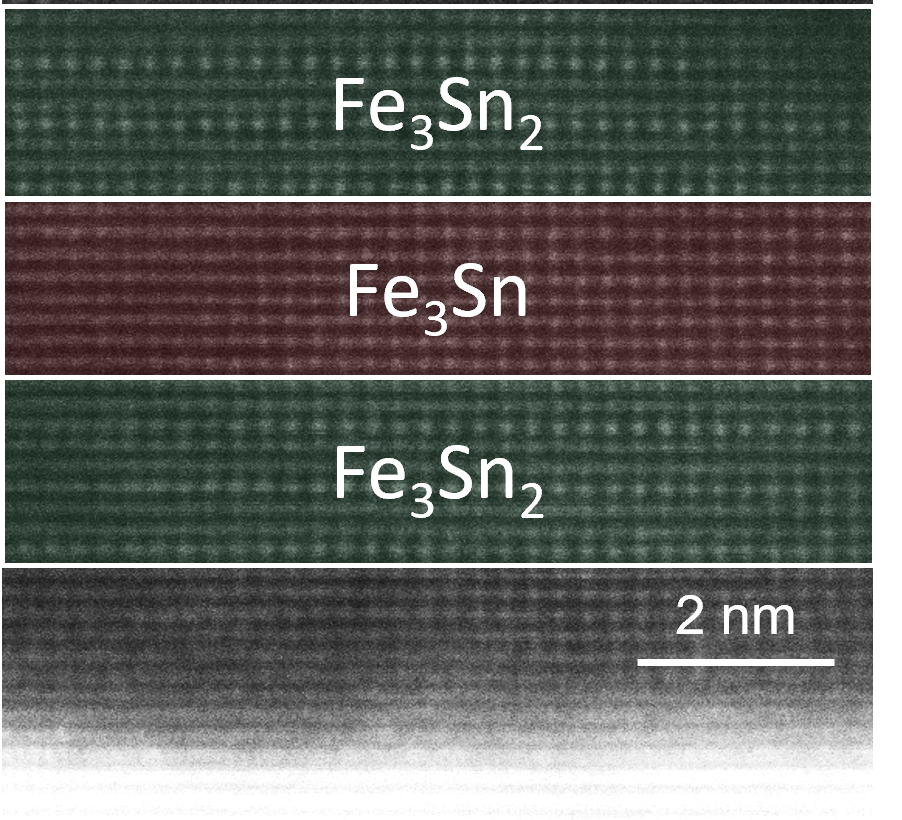}
}
   \subfloat[\label{fig:EDX_superlattice}]{
   \includegraphics[height = 1.9in]{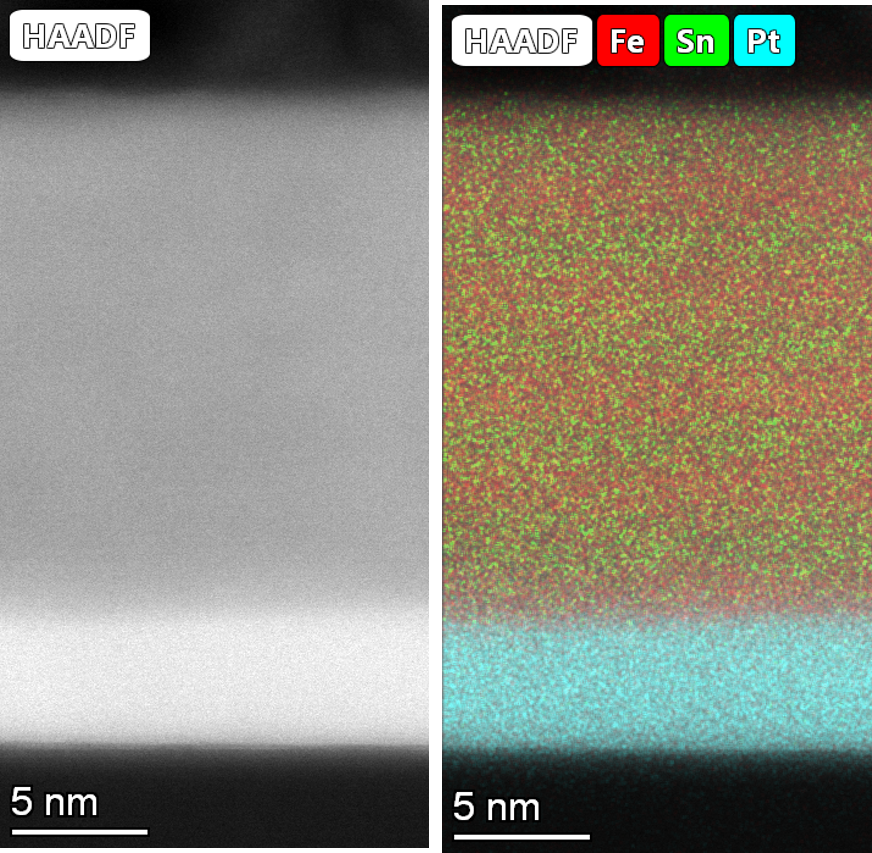}
}

  \caption{[Fe$_3$Sn$_2$/Fe$_3$Sn]$_5$ superlattice. 
  (a) RHEED pattern of  [Fe$_3$Sn$_2$/Fe$_3$Sn]$_5$ superlattice.
  (b) TEM image of [Fe$_3$Sn$_2$/Fe$_3$Sn]$_5$ superlattice. 
  (c) EDX of [Fe$_3$Sn$_2$/Fe$_3$Sn]$_5$ superlattice. 
  }
\end{figure*}

We first utilized the ANE microscope to measure a detailed hysteresis loop at a fixed position. As shown in Fig.~\ref{fig:ANE_hloop} for magnetic field along the $X$ direction, the hysteresis loop shows a gradual reversal followed by magnetization switching with coercivity of 2.3\,mT. 
This has a similar coercivity but more gradual initial reversal than the in-plane hysteresis loops obtained by MOKE (Fig.~\ref{fig:MOKE}).
The origin of the different hysteresis properties is revealed by imaging the magnetic domain structure of Fe$_3$Sn$_2$ films at a series of magnetic fields.
A representative sequence during the magnetization reversal is shown in Fig.~\ref{fig:ANE_imaging}.
Starting at -10.0\,mT, the magnetization is in a saturated state along $-X$ (blue). The reversal initiates with the nucleation of white regions with $M_x \approx 0$, mainly at the edges of the sample. 
This can be explained by the minimization of domain wall energy as the edge boundary does not contribute a domain wall energy cost. 
The nucleation at the edges initiates magnetization reversal which results in a more rounded hysteresis loop compared to the uniform films.
With increasing magnetic field, domains of opposite polarity grow inward and coalesce across the channel.
At about +1.5\,mT, the magnetic structure is in a multidomain state with characteristic features (e.g. blue and red regions) ranging from 1 to 10 microns in size.
By +2.5\,mT, most of the magnetic moments have switched to $+X$ direction, with only a few regions remaining along $-X$. 
Finally, at +10.0\,mT the magnetization reversal is complete and the films is fully saturated along $+X$.

Finally, to demonstrate the ability to control the sample structure at the atomic level, we synthesized a [Fe$_3$Sn$_2$ (2\,nm)/Fe$_3$Sn (2\,nm)]$_5$ superlattice using the AL-MBE technique.
The [Fe$_3$Sn$_2$/Fe$_3$Sn]$_5$ superlattice samples were grown under the same conditions as Fe$_3$Sn$_2$ samples but with different atomic layer deposition sequences.
For a 2\,nm Fe$_3$Sn layer, we deposit nine atomic layers of Fe$_3$Sn without any Sn$_2$ spacers. 
For a 2\,nm Fe$_3$Sn$_2$ layer, we deposit two atomic layers of Fe$_3$Sn, and one atomic layer of Sn$_2$ and repeat a total of three times.
The RHEED pattern of a [Fe$_3$Sn$_2$/Fe$_3$Sn]$_5$ superlattice along the $[11\bar{2}0]$ direction of c-sapphire is shown in Figure~\ref{fig:RHEED_superlattice}.

Such control of the stacking sequence with atomic level precision is confirmed by the HAADF STEM image of the [Fe$_3$Sn$_2$/Fe$_3$Sn] superlattice structure in Figure~\ref{fig:TEM_superlattice}. 
From the STEM image, an alternating sequence of 2\,nm Fe$_3$Sn$_2$ (false colored in green) and 2\,nm Fe$_3$Sn (false colored in red) can be observed with atomic resolution.
Within 2\,nm Fe$_3$Sn$_2$ layer, a repetitive stacking of two Fe$_3$Sn atomic layers and one Sn$_2$ atomic layer can be observed.
In contrast, we can only see Fe$_3$Sn atomic layers in 2\,nm Fe$_3$Sn layer.
The STEM-EDX chemical map further reveals the repetition of the [Fe$_3$Sn$_2$/Fe$_3$Sn]$_5$ superlattice structure along the growth direction as shown in Figure~\ref{fig:EDX_superlattice}, where a 2\,nm Fe$_3$Sn$_2$ layer has a stronger signal for the Sn element compared to a 2\,nm Fe$_3$Sn layer.

In conclusion, we report the atomic layer epitaxy growth of kagome ferromagnet Fe$_3$Sn$_2$ thin films on Pt(111)/Al$_2$O$_3$(0001) at low temperatures.
The high quality of epitaxial Fe$_3$Sn$_2$ films is confirmed by \textit{in situ} RHEED, XRD, AFM and TEM.
Low temperature growth helps to generate a sharp interface between Fe$_3$Sn$_2$ and Pt layers, which has been observed by EDX.
The magnetic properties are investigated by magneto-optical Kerr effect, SQUID magnetometry, and anomalous Nernst effect, confirming the easy-plane magnetic anisotropy of the thin films. 
Using ANE microscopy, we successfully resolve the local in-plane oriented micrometer size domains during magnetization reversal.
Finally, we demonstrate the ability to control the sample structure at the atomic level by synthesizing  [Fe$_3$Sn$_2$/Fe$_3$Sn]$_5$ superlattices and confirming their structure by TEM.
These advances enable novel heterostructures for exploring the rich physics of kagome magnets.

\section*{Acknowledgements}

S.C., B.W., I.L., N.B., D.W.M and R.K.K. acknowledge support from DARPA Grant No.~D18AP00008. A.J.B. and R.K.K. acknowledge support from AFOSR MURI 2D MAGIC Grant No.~FA9550-19-1-0390 and DOE Grant No.~DE-SC0016379. B.W. also thanks the support from Presidential Fellowship of the Ohio State University. This research was partially supported by the Center for Emergent Materials, an NSF MRSEC, under award number DMR-2011876. Electron microscopy experiments were supported by the Center for Electron Microscopy and Analysis at the Ohio State University.

\section*{Author contributions}

S.C., I.L., and R.K.K. conceived the experiments. S.C. conducted the MBE growth, AFM measurements, MOKE measurements and SQUID measurements. 
I.L. conducted the ANE measurements.
A.J.B. and I.L. conducted the XRD measurements.
B.W., N.B., and D.W.M. conducted the TEM measurements.
All authors participated in data analysis and preparation of the manuscript.

\bibliography{Fe3Sn2_and_Mn3Sn.bib}

\end{document}